# Exploiting EEG Signals for Eye Motion Tracking

R. Kovtun[1], S. Radchenko[1], A. Netreba[1], O. Sudakov[1], R. Natarov[1,2], Z. Dyka[2], I. Kabin[2] and P. Langendörfer[2,3]

[1]*Medical Radiophysics Department Taras Shevchenko National University of Kyiv,* Ukraine

[2]*IHP – Leibniz-Institut für innovative Mikroelektronik* Frankfurt (Oder), Germany

[3]*BTU Cottbus-Senftenberg* Cottbus, Germany

*Abstract*—Human eye tracking devices can help to investigate principles of processing visual information by humans. The attention focus movement during the gaze can be used for behavioural analysis of humans. In this work we describe our experimental system that we designed for synchronous recording of electroencephalographic signals, events of external tests and gaze direction. As external tests we used virtual cognitive tests. We investigated the possibility to exploit electroencephalographic signals for eye motion tracking. Our experimental system is a first step for the designing an automatic eye tracking system and can additionally be used as a laboratory equipment for teaching students.

## I. Introduction

Eye tracking and devices to detect the human's gaze direction are used to study points of selective concentration i.e. spatial and temporal focus of human attention, and for complex investigations of human behaviour [1], [2]. Some of these sophisticated studies relate to the particularities of the human gaze and visual behaviours [3]. Most eye tracking devices are used to study the visual system, in training systems, in psychology, cognitive linguistics, for evaluation of information perception or reading speed, etc. [4]-[6]. Human eye tracking technologies are now being combined with statistical methods of data processing and machine learning [7], [21] allowing to create:

- advanced driving assistance systems and humanoid visual perception in robotics [8], [9];
- systems for education and learning based on the individual characteristics of students [2], [10], [14].

Additionally, eye tracking techniques are used in medicine as communication tools for patients with specific medical/physiological conditions such as Rett syndrome or amyotrophic lateral sclerosis [11], [12].

A large number of eye tracking systems based on eye motion measurements are available. Some eye-trackers use measured mechanical motion of the eye markers (i.e. the eye markers offsets) for determining the gaze direction [13]. Other systems use optical tracking of cornea movement for that [2], [11], [12], [14]-[17]. Optical eye-tracking systems are used different tracking algorithms [18]-[20]. They can be combined with tools for analysis and classification of medical diagnostic data [21], [22]. Eye-tracking systems based on an analysis of electro myopotential distribution recorded as electrooculograms are discussed in [23].

In this research we:

- demonstrate the relationship between gaze point direction and characteristics of electroencephalographic (EEG) signals by factor analysis of the EEG data;
- investigate the possibility to use only EEG signals for eye tracking and evaluate it using an optical eye tracking system.

For demonstrating the applicability of EEG signals for determining the gaze point direction we:

- developed a special experimental system that allows to measure EEG signals in parallel to optical eye tracking as well as behaviour tracking using a virtual cognitive test [24];
- synchronized the measured data and cognitive test events applying our synchronization system based on sound events [25];
- improved and extended a previously developed method [25] for merging data from various sources to study human behavioural reactions.

The experimental system that we designed can not only be used with the virtual water maze test that we selected as a representative example for our investigations, but with many other cognitive tests [24]. We expect that our system can be extended and improved in the future so that it can be used for investigating the exchange of information between:

- human(s) and technical system(s);
- human(s) and training/teaching system(s);
- human(s) and system(s) for visual behavioural analysis.

The rest of the paper is structured as follows. In section II we describe the experimental system that we developed. In section III we discuss the applicability of EEG data for determining the gaze point i.e. we analyze EEG data and demonstrate a relation between gaze point direction and characteristics of EEG signals. We evaluate the results of the proposed approach using a blind testing i.e. in our experiments the experimenter does not have any knowledge about the gaze direction of the test person. The experimenter determines the gaze direction using the proposed approach and compares it with the original gaze direction known by the test person. The paper finishes with short conclusions.

## II. DEVELOPED EXPERIMENTAL SYSTEM

Fig. 1 shows our experimental system schematically.

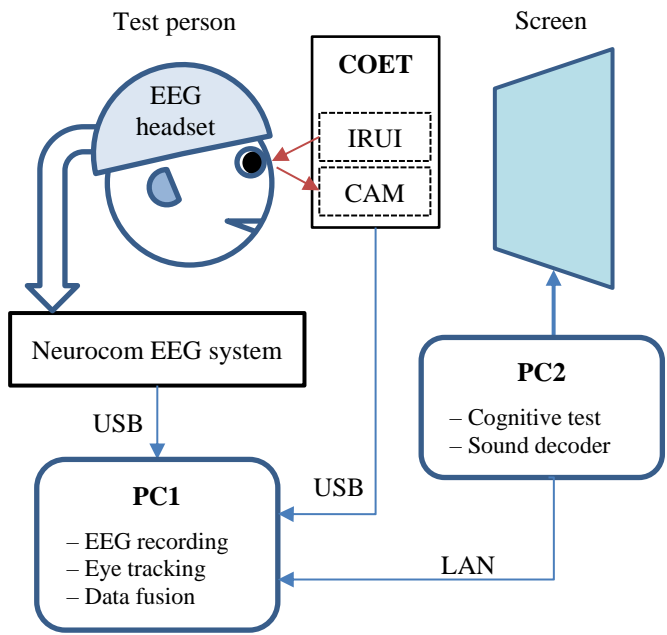

Fig. 1. Schematic representation of the developed experimental system.

In our experimental system we used two computers: PC1 and PC2. The data synchronisation between the computers is maintained by local-area network (LAN). PC1 receives data from the Neurocom EEG system and images from Contactless Optical Eyes Tracking (COET) developed by us. PC2 is used for running of the cognitive test – the virtual Morris maze test software [26], i.e. external events are generated in the maze test software. The external events are accompanied by specific sounds as we had no access to the Maze source code to register the events natively. The Maze test software uses files in a conventional sound format for playback, so we injected short sine wave sound at the beginning of each sound file. We used different frequencies for different events. The server-side of our custom-built sound decoding software, which also runs at the PC2, is able to easily capture sound from the Windows sound mixer during the walk through the Maze test, decode it in real time, and transmit corresponding events via LAN to the PC1. Computers are directly connected to each other, in order to minimize network latency. The overall delay between events and corresponding EEG signals is around 6 ms.

PC1 runs the Neurocom software to store EEG signals to the database. The client-part of our software receives events from the PC2 and writes them to the same database obeying the transmission delay.

For the eye tracking we used the third-party software solution 'EYE Writer' [12]. We calibrated this software to be able to use it with our particular camera. We tested this solution to verify the accuracy for the gaze detection. A high level of synchronization is achievable due to two additional electrocardiographic channels of the Neurocom system: we used one of them to write the event data, and the second one to store the gaze direction as a combination of two pulses of sine signals with different frequencies and durations.

Generally, the synchronization of the time and spatial information about the external events with other measured data is very important and a non-trivial data fusion task. The described data fusion technique allows to synchronously store not only events and gaze direction data with EEG data in one database, but also events from additional sources and other signals, as we did for the maze test software. The obtained data of the human behaviour should be suitable for measurement and further simultaneous analysis.

In general, our system operates as follows: the test person with a 19-chanel headset connected to the Neurocom EEG system also wears our COET device and looks at the computer screen with the Morris maze test and control the test. EEG signals with the events of the test and the gaze direction are synchronously recorded and stored to the database for the further analysis. Fig. 2 shows a test person wearing our experimental system.

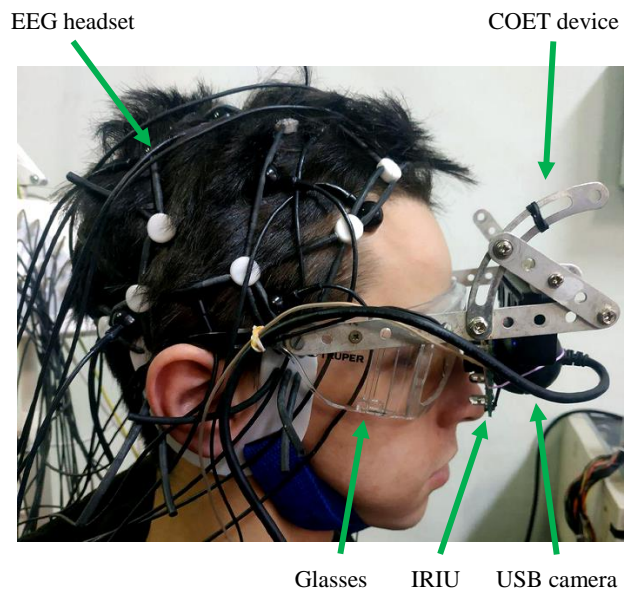

Fig. 2. Our experimental system on a test person.

Due to the inconvenience in simultaneously wearing the EEG headset and an eye-tracking device, determining the gaze direction using only EEG signals is by far more practicable, i.e. our approach can significantly decrease the complexity of evaluating and interpreting cognitive tests. This is the reason why we investigate the possibility to use only the EEG signals for determining the gaze point direction. In the rest of this section we explain some technical details of the EEG system used and the COET device developed.

### A. The COET device

The COET device uses Contactless Optical Eye Tracking principles [12], [16]. The device is capable to detect the direction of eye movements and to track it. We used common low-cost components – plastic glasses, LEDs, metal frames etc., and a high-frequency commercial camera for designing the COET device.

The camera used is a USB PlayStation EYE camera with reasonable price, a sampling rate of 120 frames per second, and an eyeglass holder. This camera, like most portable colour USB cameras, has an infrared (IR) filter blocking infrared radiation needed for correct colour representation during photography. We replaced this filter by one fully rejecting visible light but transparent for IR spectrum in order to reduce the unwanted glare from the cornea during the detection of the eye movement.

Additionally we constructed the InfraRed Illumination Unit (IRIU) that includes 8 LEDs. The uniformity of the lighting influences the gaze point determination. We selected the number and location of LEDs due to their polar radiation pattern thus the IRIU uniformly illuminates the area around the eye cornea. The modified camera and the IRIU were mounted on the eyeglasses using metal frames. We performed some adjustments of the IRIU and camera positions to achieve the best illumination and visibility of the cornea.

*B. EEG system*

The EEG headset used in our experiments is a 19-channels electroencephalographic system Neurocom. This system is a very powerful Windows PC electroencephalographic system, designed for a wide range of neurological studies (EEG recording, visual and auditory evoked potentials, video EEG, neurofeedback etc.) [27]. The Neurocom System operates with EEG signals in the range from 1 µV to 12 mV. It has a sampling frequency of 500 Hz and an effective noise value of 0.5 µV in the frequency band from 0.15 Hz to 100 Hz that covers most of the EEG oscillation bands.

The system acquires a 24 bit digital signal per channel and stores 21 data components (19 EEG channels and two additional electrocardiographic (ECG) channels) in the local computer database. The EEG system communicates with a personal computer via USB. The system is capable to perform simultaneous common-mode noise reduction to over 120 dB for all 21 components of an analogue signal.

### III. First Experiments Confirming The Concept

*A. EEG Dataset preparation*

EEG signals do not contain the direct information about the gaze direction. This information is "hidden" in the EEG signals. The shape of the measured EEG traces depends not only on the brain activity of the test person. The muscular activity of the test person can influence the shape significantly. This influence – artefacts – is a kind of noise and needs to be filtered or at least reduced before any analysis i.e. the measured EEG traces have to be prepared for the analysis. This is necessary to notice the neural processes information we need for our application.

During our experiments the test person sat and did not make any sudden movements. The main artefacts of muscular activity are caused by eyes blinking. Artefacts caused by unstable contacts of the EEG headset electrodes can be detected using the methods described in [28]. We removed the muscular activity artefacts applying the Independent Components Analysis (ICA). The ICA decomposition was performed using the approach introduced in [21] implemented in EEGLab v.15 software [29]. According to this approach the EEG data matrix *X* is presented as product of two matrices *A* and *s*. Matrix *A* contains new components that are statistically independent and orthogonal. Matrix *s* is a mixing matrix that contains contributions of EEG channels into matrix *A*. Eyes blinking and other miogram artefacts correspond to few independent components with large amplitude, low frequencies and large contributions from electrodes close to eyes end neck. These components are removed from matrix *A* and then EEG signals are reconstructed. Fig. 3 shows a part of the measured EEG traces with artefacts. Fig. 4 shows the same part of the EEG traces after being processed using ICA i.e. muscular activity artefacts were successfully removed. The EEG electrodes were located on the scalp of the test person corresponding to the international 10-20 System of Electrode Placement. All EEG channels are listed on the right side in Fig. 3 and Fig. 4.

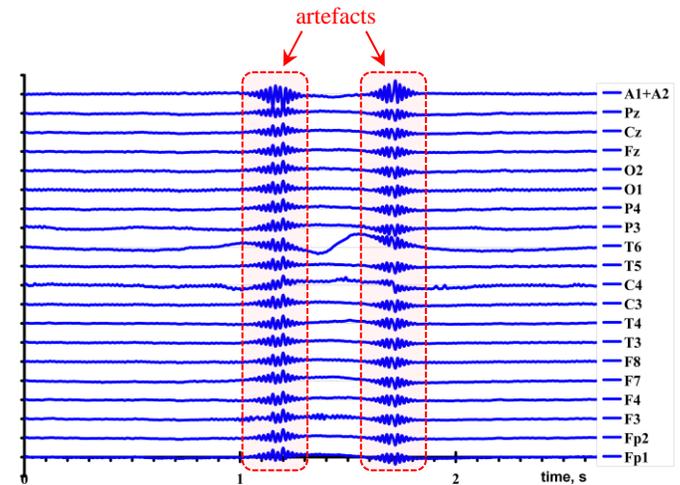

Fig. 3. Muscular activity artefacts in the measured EEG signal traces.

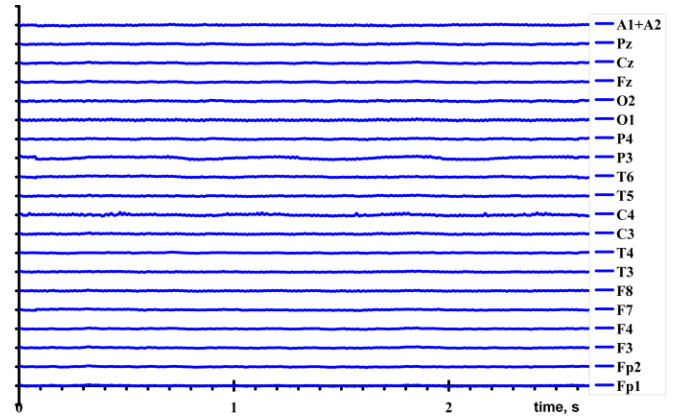

Fig. 4. Result of pre-processing the traces using PCA: muscular activity artefacts are successfully removed.

*B. Electroencephalograms Factor Analysis*

Our analysis of visual activity of a test person exploits the assumption that the EEG signals recorded synchronously with the gaze point should have common features. The models that connect EEG signals' features with gaze direction are unknown. In this work we approximated this relation using a linear model. EEG signals were processed by exploratory factor analysis to determine common components [30].

Factor analysis was performed for electroencephalograms corresponding to different positions of the gaze point on the screen. Our first step was measuring EEGs for training. For this purpose we used a specifically prepared image (Fig. 5), divided onto four equal quadrants. We recorded EEG data from three male test persons (subjects) about 20 years old. Each subject was asked to look at the center of each quadrant during a certain time (dots represents the point of gaze).

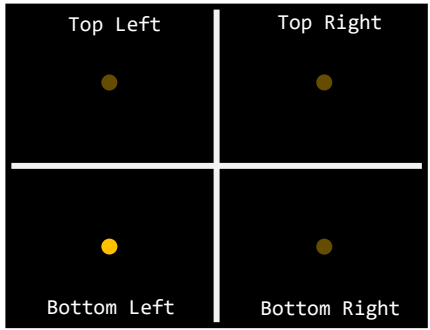

Fig. 5. Training image: test person looked at a centre of each quadrant in a predefined sequence.

We differentiate four diagonal gaze point directions (further also coordinates) only: top left, bottom left, top right and bottom right. Gaze point coordinates were determined using the COET device and the EYE Writer software. For each diagonal gaze point direction we measured 19 channel EEG signals during about 1 minute. We represented the 19 parallel measured signals as one signal i.e. we placed the signals from each channel serially after each other. Thus, we obtained four 19 minutes long traces. Each trace, i.e. the set of the feature vectors, corresponds to one gaze point direction. These sets of feature vectors for 3 test persons were analysed with IBM SPSS Statistics v.22 [31]. We applied Principal components analysis approach and Kaiser criterion. Fig. 6 shows the components' eigenvalues calculated for each test person.

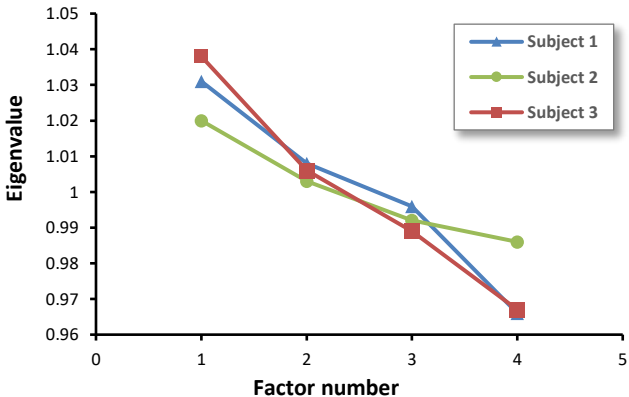

Fig. 6. Latent components eigenvalues for three test persons.

There are only two factors with eigenvalues higher than 1 (see Fig. 6). Thus, we concentrated on these two components only. The averaged scores of these components for different gaze point directions are shown in Fig. 7.

The data in Fig. 7 confirm our assumption that spatial and temporal dependences of EEG signals are associated with subjects' actions and different spots of human attention. The directions of gaze points are well defined by main components' eigenvalues. We used this fact for determining the "unknown" gaze point direction.

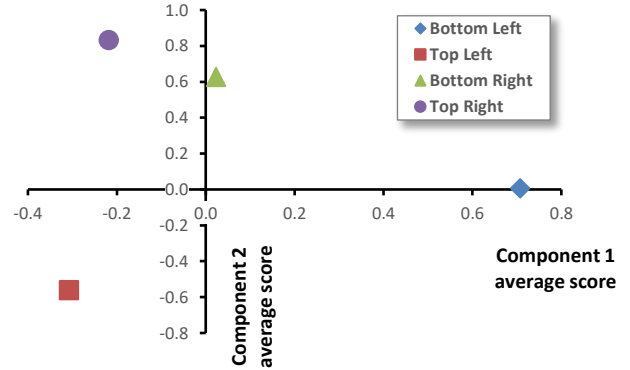

Fig. 7. Averaged scores of two main components for 4 gaze point directions.

## C. Blind test experiments

To confirm the possibility to determine the gaze point direction by EEG data only we performed two experiments. Only one test person – Subject 3 – participated in these experiments. EEG signals were measured during 1 minute for each of two gaze point directions that were chosen by the test person himself. The first gaze direction was chosen to be identical to one of the indicated training directions (experiment 1). The second gaze direction was chosen different from the indicated training directions (experiment 2). It was expected, that values of components scores will be similar to the training data in case of identical gaze directions. The values of component scores obtained for the both experiments are shown in Table I.

TABLE I. COMPONENT SCORES CALCULATED FOR THE BLIND TEST EXPERIMENTS

| Experiment | Component scores | |
|---|---|---|
| | *for component 1* | *for component 2* |
| 1 | 0.014 | 0.682 |
| 2 | 0.475 | –0.42 |

In experiment 1 the values of component scores are close to the component scores for the gaze point direction "bottom right". The test person as well as the eye tracking software confirmed the correctness of the obtained result.

The values of component scores calculated for the experiment 2 differ significantly from the data obtained for all 4 training gaze directions. Corresponding to information from the test person the chosen gaze point direction was "bottom" i.e. didn't coincide with any of training directions.

## IV. CONCLUSION

Analysis of human behaviour using virtual cognitive tests is a basis for training and teaching systems. Determining the gaze point direction is a part of such systems. Synchronizing the events of the cognitive tests, gaze point direction of a test person and his/her reaction is a non-trivial task. In this work we presented an experimental system that we developed for human behaviour analysis. Our experimental system allows to measure EEG signals in parallel to optical eye tracking as well as behaviour tracking.

While designing our experimental system we assumed that EEG signals can contain information about the gaze point direction. Determining the gaze direction using EEG signals only is by far more practicable: the complexity of interpreting cognitive tests can be significantly decreased using EEG signals i.e. without any eye tracking device and software for its synchronization with the EEG signals. We performed here only first experiments for confirming our assumption i.e. we demonstrated the relationship between gaze point direction and characteristics of electroencephalographic signals using factor analysis of the EEG data.

Accuracy and precision of determining the gaze point direction are important parameters. We will improve the experiments described here in our future work: the duration of a look at each gaze point for each test person has to be decreased whereas the number of gaze points for each test person as well as the number of test persons have to be increased significantly.